\documentstyle[aps,preprint]{revtex}

\begin{document}
\title{Feshbach resonances in a quasi-2D atomic gas}
\author{M. Wouters, J. Tempere$^{\ast }$, J. T. Devreese$^{\ast \ast }$}
\address{TFVS, Universiteit Antwerpen, Universiteitsplein 1, 2610 Antwerpen, Belgium. }
\date{May 11th, 2003}
\maketitle

\begin{abstract}
Strongly confining an ultracold atomic gas in one direction to create a
quasi-2D system alters the scattering properties of this gas. We investigate
the effects of confinement on Feshbach scattering resonances and show that
strong confinement results in a shift in the position of the Feshbach
resonance as a function of the magnetic field. This shift, as well as the
change of the width of the resonance, are computed. We find that the
resonance is strongly damped in the thermal gas, but in the condensate the
resonance remains sharp due to many-body effects. We introduce a 2D model
system, suited for the study of resonant superfluidity, and having the same
scattering properties as the tightly confined real system near a Feshbach
resonance. Exact relations are derived between measurable quantities and the
model parameters.
\end{abstract}


\section{Introduction}

\tighten The ability to tune the interatomic interaction strength by using a
Feshbach resonance has led to renewed interest in the scattering theory of
cold trapped atoms, in particular in relation with the realization of
resonant superfluidity of fermionic atoms \cite{OharaSCI298}. Also the
ability to confine Bose-Einstein condensates in periodic potentials has
attracted much interest recently, since the resulting optical lattices are
ideally suited to probe fundamental quantum effects such as the superfluid
to Mott insulator transition \cite{GreinerNAT}. One-dimensional optical
lattices allow to create stacks of `pancake'-shaped clouds filled with
neutral atoms \cite{inguscioScience293,burgerEPL57,FortPRL90}, where an
effectively two-dimensional (2D) regime can be achieved. In this Letter, we
address a central question on the crossroads between the two aforementioned
achievements, Feshbach scattering and optical lattices: what happens to the
Feshbach resonance as the atoms are strongly confined and become (quasi-)
two-dimensional ?

In Sec. II, we derive the 2D resonant scattering amplitude in terms of
quantities that can be measured in experiments on three-dimensional (3D)
gases. For many-body calculations, it can be helpful to replace the complex
real atomic interaction by a simpler effective interaction. To achieve this,
we investigate in Sec. III a schematic model of coupled atoms and molecules
that interact through a contact potential. Such model systems already have
been used before in 3D \cite{kokkelmansPRA65}, but the connection between
the model parameters and the experimental quantities in 2D is not yet
available. By matching the scattering properties of this model to the
results from Sec. II, we derive expressions for the strengths of the
effective{\bf \ }contact interactions as a function of experimentally
measurable (and tunable) quantities. This solves the problem of determining
the parameters of a Friedberg-Lee like Hamiltonian \cite{FriedbergLeePRB40}
for atomic gases. Such models have been studied in the context of high
temperature superconductivity, but for these systems fixing the model
parameters remained a problem.

\section{Quasi 2D resonant scattering}

The existence of a Feshbach resonance relies on the presence of\ a bound
state in a closed channel, with an energy equal to the\ energy of the
colliding particles. In the case of trapped cold atoms, the closed channels
involve states with higher Zeeman energy in the external magnetic field. The%
{\bf \ }energy of the bound state (molecule) is tunable by changing this
magnetic field. Because these molecular states are small compared to\ the
length scale of the confinement potential, they are not expected to be
influenced strongly by the trapping potential. However, this is only a
simplified picture. Because of the coupling between the molecular state and
the open channels, the binding energy of the molecule acquires a complex
shift (a change of the binding energy and a decay width). This shift depends
on the nature of the open channels and consequently on the tight confinement.

If the external confinement potential varies slowly\ on the scale of the
atomic interaction range, the two-body\ scattering can be treated
analytically, even in the vicinity of a Fesh\-bach resonance (see also Ref. 
\cite{juliennePRA66}). In the following, we denote the range of the van der
Waals interatomic potential with $R_{e}$ and the oscillator length of the
tight confinement in the axial direction by $\ell _{z}=\sqrt{\hbar /\left(
2m_{\text{red}}\omega _{z}\right) },$ with $m_{\text{red}}=\left(
m_{1}^{-1}+m_{2}^{-1}\right) ^{-1}$ the reduced mass of the scattering atoms
and{\bf \ }$\omega _{z}${\bf \ }the characteristic frequency of the trap in
the axial direction. For interatomic distances $R_{e}\ll r\ll \ell _{z},$
both the interatomic and the confinement potential\ are negligible compared
to the total energy and the asymptotic $s$-wave scattering wave function is
consequently that for free 3D motion 
\begin{equation}
\psi \left( {\bf r}\right) \propto \frac{\sin \left( qr+\delta _{3D}\left(
E\right) \right) }{qr},  \label{psi3D}
\end{equation}
with $q=\sqrt{2m_{\text{red}}E}/\hbar $. The energy $E$\ is measured in the
center of mass frame of the scattering atoms, relative to the energy of two
atoms in the open channel at infinity. The phase shift $\delta _{3D}\left(
E\right) $ is determined by the solution of the free 3D scattering problem.
At a Feshbach resonance, this phase shift shows resonant behavior, but the
wave function (\ref{psi3D}) remains valid. Close to the resonance, the
energy dependence of $\delta _{3D}\left( E\right) $ is important, because it
characterizes the resonance. From the analytic theory of Feshbach resonances 
\cite{moerdijkPRA51} it follows that it is given by 
\begin{equation}
\delta _{3D}\left( E\right) =\delta _{bg}-\arctan \left[ \frac{\Gamma _{F}/2%
}{E-E_{F}-\Delta _{F}}\right] ,  \label{delta3D}
\end{equation}
where $\delta _{bg}${\bf \ }is the 3D `background' scattering phase shift
related only to the open channels, and{\bf \ }$\Gamma _{F},E_{F}$ and $%
\Delta _{F}$ are respectively the width and{\bf \ }the position of the 3D
Feshbach{\bf \ }resonance and the shift of that position due to the coupling
with the open scattering channels.

Next, we look at large interatomic distances $r\gg R_{e}$, where the motion
is governed by the\ confining potential and must be a solution of 
\begin{equation}
\left[ -\frac{\hbar ^{2}}{2m_{\text{red}}}\nabla ^{2}+\frac{1}{2}m_{\text{red%
}}\omega _{z}^{2}z^{2}\right] \psi \left( {\bf r}\right) =E\psi \left( {\bf r%
}\right) .  \label{eqquasi2D}
\end{equation}
Petrov and Shlyapnikov have analyzed this problem \cite{petrovPRA64}. Using
their solution of Equation (\ref{eqquasi2D}) and matching it for $%
r\rightarrow 0$ to (\ref{psi3D}) we find for the scattering amplitude 
\begin{equation}
f_{2D}\left( E\right) =\frac{2\sqrt{2\pi }}{-q\ell _{z}\cot \delta
_{3D}\left( E\right) +%
{\textstyle{1 \over \sqrt{2\pi }}}%
\left[ \ln \left( 
{\textstyle{C\hbar \omega _{z} \over \pi E}}%
\right) +i\pi \right] },  \label{f}
\end{equation}
where $\delta _{3D}\left( E\right) $ is given by (\ref{delta3D}) and $%
C=0.915 $ \cite{petrovPRA64}. Making use of formulas derived in \cite
{moerdijkPRA51}, $\cot \delta _{3D}\left( E\right) ${\bf \ }can be related
to measurable quantities 
\begin{equation}
\cot \delta _{3D}\left( E\right) =-%
{\displaystyle{1 \over qa_{bg}}}%
\frac{E-\left[ B-B_{0}+\Delta B(qa_{bg})^{2}\right] \Delta \mu }{E-\left(
B-B_{0}-\Delta B\right) \Delta \mu },  \label{cotd3d-bis}
\end{equation}
where $B_{0}$ and $\Delta B$ are respectively the 3D experimental position
and width of the Feshbach resonance, and $a_{{\sl bg}}$\ is the background
scattering length. $\Delta \mu =\mu _{1}+\mu _{2}-\mu _{res}$ ; $\mu _{1,2}$
are the atomic magnetic moments and $\mu _{res}$ is the magnetic moment of
the resonant state.

Formula (\ref{cotd3d-bis}), together with (\ref{f}) gives the 2D resonant
scattering amplitude for a quasi 2D situation in terms of parameters that
can be measured in 3D experiments. It should be noted that for low energies
(small compared to $E_{F}+\Delta _{F},$ which is typically of the order $\mu
_{B}\times 1$ G $=k_{B}\times 67.17$ $\mu $K, where $\mu _{B}$ is the Bohr
magneton), the energy dependence of (\ref{cotd3d-bis}) is weak and $\cot
\delta _{3D}\left( E\right) $ is well approximated by its value for $E=0,$
namely $-q\ell _{z}\cot \delta _{3D}\left( E\right) \rightarrow \ell
_{z}/a_{3D},$ which\ would yield the usual relation $a_{3D}=a_{bg}[1-\Delta
B/(B-B_{0})]$ for a Feshbach resonance in 3D.

A remarkable feature of the expression (\ref{f}) is that it shows that in
two dimensions the scattering amplitude never diverges, because $\left|
f_{2D}\left( E\right) \right| ^{2}\leq 16$, irrespective of $E$ or $E_{F}$.
This is in contrast with the 3D case where the imaginary part in the
denominator of the scattering amplitude ($=\Gamma _{F}$) is proportional
with $\sqrt{E}$ and vanishes as $E\rightarrow 0$ \cite{moerdijkPRA51}$.$
However, when many-body effects are taken into account for bosonic atoms,
the damped character of the Feshbach resonance in two dimensions disappears.
Lee {\it et al.}{\em \ }show in Ref. \cite{leePRA65} that in the limit of
zero temperature in a BEC the relevant matrix element of the 2D many-body $T$%
-matrix $T_{MB}$ is to a good approximation equal to the analytic
continuation of the two-body $T$-matrix $T_{2b}\left( E\right) =\left( \hbar
^{2}/m\right) f_{2D}\left( E\right) $, given by $\left\langle 0\right|
T_{MB}\left( E=0\right) \left| 0\right\rangle =\left\langle 0\right|
T_{2b}\left( -\lambda \right) \left| 0\right\rangle ,$ where $\lambda $ is
the chemical potential of the Bose-gas. By this procedure, the $i\pi $\ term
in expression (\ref{f}) disappears, so that a sharp resonance is expected in
a Bose-Einstein condensed gas at low temperatures. Formulas (\ref{f}) and (%
\ref{cotd3d-bis}), imply that the resonance position in the quasi-2D
condensate, $B_{0}^{2D}$, is different from $B_{0}.$ The shift $%
B_{0}^{2D}-B_{0}$ depends on the chemical potential and diverges for $%
\lambda =\varepsilon ^{\ast }\left( \omega _{z}\right) =C\hbar \omega
_{z}/\pi \exp [\sqrt{2\pi }\ell _{z}/a_{bg}]$, which is the energy of the
weakly bound 2D state if only the background scattering is taken into
account \cite{petrovPRA64}. At this value of the scattering energy, also the
width of the resonance in a BEC diverges. For the thermal gas, there are no
coherent many-body effects and $\left| f_{2D}\left( E\right) \right| ^{2}$
(expression (\ref{f})) is the relevant quantity, for example in the
calculation of the rethermalization rates.

The scattering amplitude in the neighborhood of the Feshbach resonance at $%
B_{0}=154.9$ G in a $^{85}$Rb condensate (experimentally explored in Ref. 
\cite{donleyPRA64}) is shown in Fig. 1 as a function of the confinement
frequency and the magnetic field. The resonance parameters are $\Delta
B=11.9 $ G$,$ $a_{bg}=-380a_{0}$, $\Delta \mu =-2.23\mu _{B}$ \cite
{donleyPRA64,KokkelmansPRL89}. The chemical potential was fixed at $\lambda
/\hbar =5$ kHz. The white regions in Fig. 1 correspond to $|f(-\lambda
)|>100 $ and indicate the location of the resonance. The divergence of the
shift at $\varepsilon ^{\ast }\left( \omega _{z}\right) =\lambda $ can
clearly be seen. Inset (b) of Fig. 1 shows the scattering amplitude $%
f_{2D}(-\lambda )$ for condensate atoms at $\omega _{z}=0.4$ MHz, as a
function of magnetic field. For this value of $\omega _{z}$, we find a shift
of $B_{0}^{2D}-B_{0}=-16.5$ G, indicated by an arrow in inset (b).

In experiments, it is the radial confinement and the number of particles
that are controlled, rather than the chemical potential. The chemical
potential for a homogeneous system is equal to $\lambda =\left\langle
0\right| T_{MB}\left( E=0\right) \left| 0\right\rangle \times n_{b},$ with $%
n_{b}$ the density of bosons and accordingly has to be determined
self-consistently. When the chemical potential is determined
self-consistently in the Thomas-Fermi approximation, again we find that the
resonance is shifted. But to the left of the resonance no self-consistent
solution for $\lambda $ can be found. In this region we expect the
interactions to become attractive, and the absence of a self-consistent
solution can be linked to the instability occurring in condensates with
attractive interactions.

In Fig. 2, $\left| f_{2D}\left( E\right) \right| ^{2}$, the squared modulus
of the scattering amplitude for the thermal cloud, is plotted for $%
E/k_{B}=100$ nK. For the thermal gas the maximum value of the scattering
amplitude is bounded.

\section{2D contact scattering model}

To avoid having to take into account the details of the real interaction, it
is convenient to introduce a 2D model with an effective contact potential
that has the same scattering properties as the real interatomic potential in
the quasi-2D system. This effective potential can then be used to tackle
many-body problems such as the superfluidity of a two-component fermion gas.
Kokkelmans {\it et al.} have performed such a study for the 3D case \cite
{kokkelmansPRA65}. They introduced a molecular degree of freedom and showed
that the model of atoms and molecules that are converted into each other
leads to a scattering length of the Feshbach type. We construct an analogous
model to describe resonant scattering in a quasi-2D gas. Our model 2D
Hamiltonian for the scattering of two different particles $1$ and $2$
(bosons or fermions)\ takes the form 
\begin{eqnarray}
\hat{H} &=&-%
{\displaystyle{\hbar ^{2} \over 2}}%
\int d^{2}{\bf r}\,\left[ \hat{\psi}_{1}^{+}\left( {\bf r}\right) \frac{%
\nabla ^{2}}{m_{1}}\hat{\psi}_{1}\left( {\bf r}\right) +\hat{\psi}%
_{2}^{+}\left( {\bf r}\right) \frac{\nabla ^{2}}{m_{2}}\hat{\psi}_{2}\left( 
{\bf r}\right) \right] +V^{P}\int d^{2}{\bf r}\text{ }\hat{\psi}%
_{1}^{+}\left( {\bf r}\right) \hat{\psi}_{2}^{+}\left( {\bf r}\right) \hat{%
\psi}_{2}\left( {\bf r}\right) \hat{\psi}_{1}\left( {\bf r}\right)  \nonumber
\\
&&+\int d^{2}{\bf r}\text{ }\hat{\psi}_{m}^{+}\left( {\bf r}\right) \left( 
\frac{-\hbar ^{2}\nabla ^{2}}{2m_{m}}+\nu \right) \hat{\psi}_{m}\left( {\bf r%
}\right) +\left[ g_{m}\int d^{2}{\bf r}\text{ }\hat{\psi}_{m}^{+}\left( {\bf %
r}\right) \hat{\psi}_{2}\left( {\bf r}\right) \hat{\psi}_{1}\left( {\bf r}%
\right) +h.c\right] ,  \label{Hammie}
\end{eqnarray}
with $\nu $ the binding energy of the molecule and $g_{m}$\ is the parameter
related to the formation of a molecule out of two atoms. $V^{P}$ is the
interaction far from resonance and $\hat{\psi}_{1},\hat{\psi}_{2}$ and $\hat{%
\psi}_{m}$ annihilate respectively the atoms $1$, $2$ and molecules. We
solve the scattering problem for two atoms in the center of mass frame along
the lines of Ref. \cite{kokkelmansPRA65} and we find for the scattering
amplitude 
\begin{equation}
f_{2D}\left( {\bf k},{\bf p}\right) =2\sqrt{2\pi }\left\{ \frac{2\sqrt{2\pi }%
\left( \frac{\hbar ^{2}k^{2}}{2m_{\text{red}}}-\nu \right) }{\frac{2m_{\text{%
red}}}{\hbar ^{2}}V^{P}\left( \frac{\hbar ^{2}k^{2}}{2m_{\text{red}}}-\nu
\right) +\left| g_{m}\right| ^{2}\frac{2m_{\text{red}}}{\hbar ^{2}}}+%
{\displaystyle{1 \over \sqrt{2\pi }}}%
\left( \ln \left[ 
{\displaystyle{K_{c}^{2} \over k^{2}}}%
-1\right] +i\pi \right) \right\} ^{-1},  \label{fmod}
\end{equation}
which is independent of ${\bf p}$. $K_{c}$ is a momentum cutoff, introduced
to keep the integrals over ${\bf p}$ finite. Expression (\ref{fmod}) has the
same structure as (\ref{f}). Comparing both formulas, using (\ref{cotd3d-bis}%
), gives the relation between the physical quantities and the model
parameters. With the notation $\chi =2\Delta B\Delta \mu a_{bg}^{2}m_{\text{%
red}}/\hbar ^{2}$, we find 
\begin{equation}
V^{P}=\frac{1}{1-\chi }\frac{\sqrt{2\pi }\hbar ^{2}}{m_{\text{red}}}\frac{%
a_{bg}}{\ell _{z}}  \label{VP}
\end{equation}
for the non-resonant interaction. The molecular binding{\bf \ }energy is 
\begin{equation}
\nu =\left( B-B_{0}\right) \Delta \mu /(1-\chi )  \label{nu}
\end{equation}
and the coupling strength is given by 
\begin{equation}
\left| g_{m}\right| ^{2}=V^{P}\left( \Delta B\Delta \mu +\chi \nu \right) .
\label{gm}
\end{equation}
From the comparison of (\ref{f}) and (\ref{fmod}), it follows that the
cutoff momentum $K_{c}$ is related to the tight confinement harmonic
oscillator length by $K_{c}=\sqrt{C/(\pi \ell _{z}^{2})}.$ The right hand
side of expression (\ref{gm}) can be negative for some values of $\chi $ and 
$B.$ Because the description in terms of atoms and molecules describes the
physics close to the resonance, we need certainly that for $B=B_{0}$ the RHS
of (\ref{gm}) is positive. From the theory of reference \cite{moerdijkPRA51}
it follows that $a_{bg}\Delta B\Delta \mu >0,$ so our model (\ref{Hammie})
is applicable only for $\chi <1$. For the case of $^{85}$Rb around $%
B_{0}=154.9$ G, we find that $\chi =-126.3<1$\ so that the model is
applicable. In contrast with the 3D case, the parameters $V^{P}$ and $g_{m}$
in 2D do not depend on the cutoff momentum and need no renormalization, but
the cutoff momentum itself is fixed and depends on the tight confinement
frequency. Thus, formulas (\ref{VP}),(\ref{nu}) and\ (\ref{gm}), together
with the knowledge of the experimental quantities $a_{bg}$, $B_{0}$, $\Delta
B$\ and $\Delta \mu $\ from the 3D Feshbach resonance, fix unambiguously the
parameters of the model Hamiltonian (\ref{Hammie}). This allows us to
describe a quasi-2D mixture of gases that are tightly confined in the axial
direction by a harmonic potential with characteristic length $\ell _{z}$. As
a first check of the model we have found that far from resonance this
contact model with a finite cut off reproduces the critical temperature for
superfluidity derived in Ref. \cite{petrovPRA67}.

\section{Conclusions}

The scattering of cold neutral atoms that are strongly confined in one
spatial direction is studied analytically close to a Feshbach resonance. We
find that there exists a confinement-induced shift in the position of the
Feshbach resonance as a function of the magnetic field, \ and show that for
a $^{85}$Rb condensate this shift is experimentally detectable. We also find
that the Feshbach resonance in the confined thermal gas is damped, in the
sense that the scattering amplitude cannot diverge. Based on the present
analysis of Feshbach scattering in a quasi-2D gas, we set up a Friedberg-Lee
like model system with an effective contact interaction, suitable for the
study of resonant superfluidity in confined gases. Exact relations are
derived linking the model parameters to experimentally measurable quantities.

Two of the authors (M. W. and J. T.) are supported financially by the Fund
for Scientific Research - Flanders (Fonds voor Wetenschappelijk Onderzoek --
Vlaanderen). This research has been supported financially by the FWO-V
projects Nos. G.0435.03, G.0306.00, the W.O.G. project WO.025.99N.and the
GOA BOF UA 2000, IUAP.

\begin{figure}[tbp]
\caption{The scattering amplitude $f_{2D}\left( -\protect\lambda \right) $
for a quasi-2D $^{85}$Rb {\it condensate}, with chemical potential $\protect%
\lambda /\hbar =5$ kHz, is shown color-coded as a function of the magnetic
field and the trapping frequency in the tightly confined direction $\protect%
\omega _{z}$, close to the 3D Feshbach resonance at $B_{0}=154.9$ G. The
white regions correspond to $|f(-\protect\lambda )|>100$ and contain the
location of the resonance in the quasi-2D condensate. In the insets (a) and
(b) $f_{2D}\left( -\protect\lambda \right) $ is plotted at fixed magnetic
field and fixed frequency, respectively, as indicated by the lines (a) and
(b). At a confinement frequency of $\protect\omega _{z}=0.4$ MHz (inset
(b)), the Feshbach resonance is shifted by $-16.5$ G, as shown by the arrow. 
}
\label{figure3}
\end{figure}
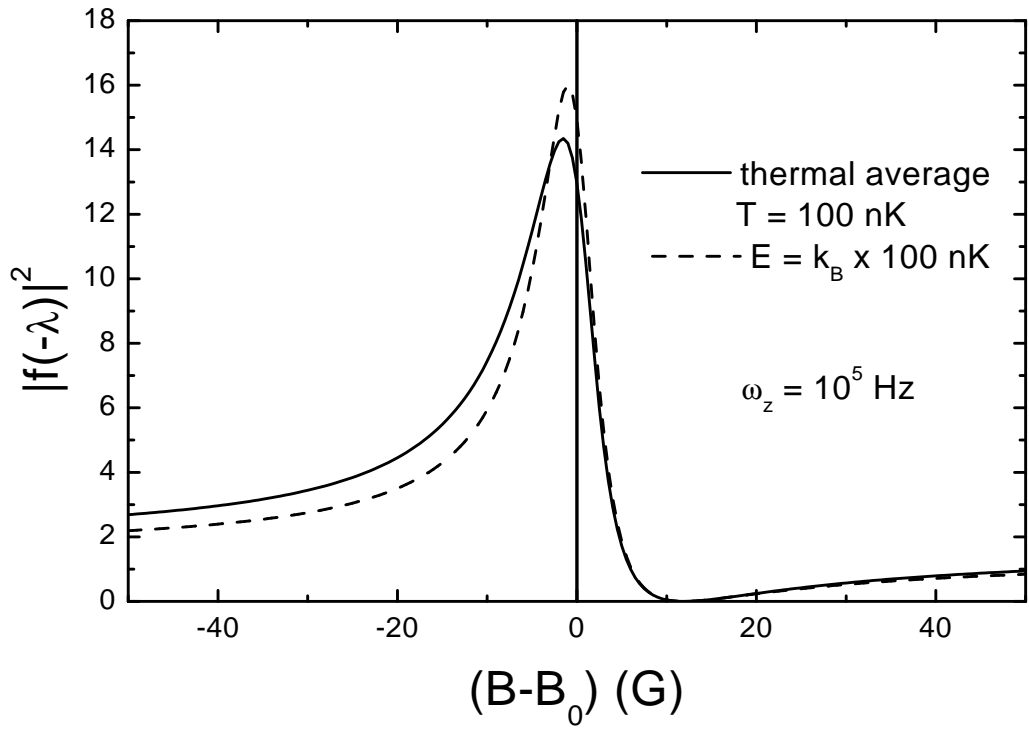
\begin{figure}[tbp]
\caption{The modulus square of the scattering amplitude $\left\vert
f(E)\right\vert ^{2}$ for a quasi-2D $^{85}$Rb {\it thermal cloud} is shown
for $T=100$ nK close to the Feshbach resonance at $B_{0}=154.9$ G for and $%
\protect\omega _{z}=10^{5}$ Hz. Results are shown with (full line) and
without (dashed line) taking thermal broadening into account. In both cases,
the divergence of the scattering amplitude has disappeared, due to the 2D
character of the gas. }
\label{figure4}
\end{figure}


\begin{references}
\bibitem[*]{A1}  Also at: Lyman Laboratory of Physics, Harvard University,
Cambridge MA02138.

\bibitem[{*}*]{A2}  Also at: T.U.E., P.B.513, 5600MB Eindhoven, The
Netherlands.

\bibitem{OharaSCI298}  M. Holland {\it et al.} Phys. Rev. Lett. {\bf 87},
120406 (2001); O'Hara {\it et al.}, Science {\bf 298}, 2179 (2002).

\bibitem{GreinerNAT}  M. Greiner {\it et al.}, Nature (London) {\bf 415}, 39
(2002).

\bibitem{inguscioScience293}  F. S. Cataliotti {\it et. al.}, Science {\bf %
293}, 843 (2001).

\bibitem{burgerEPL57}  S. Burger {\it et. al., }Europhys. Lett. {\bf 57}, 1
(2002).

\bibitem{FortPRL90}  C. Fort {\it et. al., }Phys. Rev. Lett. {\bf 90},
140405 (2003).

\bibitem{kokkelmansPRA65}  S. Kokkelmans {\it et. al.,} Phys. Rev. A {\bf 65}%
, 053617 (2002).

\bibitem{FriedbergLeePRB40}  R. Friedberg and T. D. Lee, Phys. Rev. B {\bf 40%
}, 6745 (1989).

\bibitem{juliennePRA66}  E. L. Bolda, E. Tiesinga, P.S. Julienne, Phys. Rev.
A {\bf 66}, 013403 (2002).

\bibitem{moerdijkPRA51}  A.J. Moerdijk, B.J. Verhaar and A. Axelsson, Phys.
Rev. A {\bf 51}, 4852 (1995).

\bibitem{petrovPRA64}  D.S. Petrov and G.V. Shlyapnikov, Phys. Rev. A {\bf 64%
}, 012706 (2001).

\bibitem{leePRA65}  M.D. Lee {\it et al.}, Phys. Rev. A {\bf 65}, 043617
(2002).

\bibitem{donleyPRA64}  J.L. Roberts {\it et al.}, Phys. Rev. A {\bf 64},
024702 (2001).

\bibitem{KokkelmansPRL89}  S. Kokkelmans and M. J. Holland, Phys. Rev. Lett. 
{\bf 89}, 180401 (2002).

\bibitem{petrovPRA67}  D. S. Petrov, M. A. Baranov, and G. V. Shlyapnikov,
Phys. Rev. A {\bf 67}, 031601 (2003).
\end{references}
\end{document}